\documentclass{emulateapj}
\usepackage{psfig}

\begin{document}

\title{Spitzer Observations of Centaurus A: Infrared Synchrotron Emission from  the Northern Lobe}



\author{M.~H. Brookes$^1$, C.~R. Lawrence$^1$, J. Keene$^1$, D. Stern$^1$, V. Gorijan$^1$, M. Werner$^1$}\altaffilmark{1}
\altaffiltext{1}{Jet Propulsion Laboratory, California Institute of Technology,
    Pasadena, CA 91109}
\author{V. Charmandaris$^{2,3}$}
\altaffiltext{2}{Department of Physics, University of Crete,   
GR-71003, Heraklion, Greece}
\altaffiltext{3}{Astronomy Department, Cornell University,Space Sciences Building, Ithaca, NY 14853-6801, USA}

\email{Mairi.H.Brookes@jpl.nasa.gov}

\def\squig{$\sim\!\!$}
\def\arcs{\ifmmode {^{\scriptscriptstyle\prime\prime}}
          \else $^{\scriptscriptstyle\prime\prime}$\fi}
\def\arcm{\ifmmode {^{\scriptscriptstyle\prime}}
          \else $^{\scriptscriptstyle\prime}$\fi}
\def\microns{\ifmmode \,\mu$m$\else \,$\mu$m\fi}
\def\micron{\microns}
\def\deg{\ifmmode^\circ\else$^\circ$\fi}
\def\pdeg{\ifmmode $\setbox0=\hbox{$^{\circ}$}\rlap{\hskip.11\wd0 .}$^{\circ}
          \else \setbox0=\hbox{$^{\circ}$}\rlap{\hskip.11\wd0 .}$^{\circ}$\fi}
\def\arcs{\ifmmode {^{\scriptscriptstyle\prime\prime}}
          \else $^{\scriptscriptstyle\prime\prime}$\fi}
\def\arcm{\ifmmode {^{\scriptscriptstyle\prime}}
          \else $^{\scriptscriptstyle\prime}$\fi}
\newdimen\sa  \newdimen\sb
\def\parcs{\sa=.07em \sb=.03em
     \ifmmode \hbox{\rlap{.}}^{\scriptscriptstyle\prime\kern -
\sb\prime}\hbox{\kern -\sa}
     \else \rlap{.}$^{\scriptscriptstyle\prime\kern -\sb\prime}$\kern -\sa\fi}
\def\parcm{\sa=.08em \sb=.03em
     \ifmmode \hbox{\rlap{.}\kern\sa}^{\scriptscriptstyle\prime}\hbox{\kern-\sb}
     \else \rlap{.}\kern\sa$^{\scriptscriptstyle\prime}$\kern-\sb\fi}

\begin{abstract}
We present measurements obtained with the {\it Spitzer Space Telescope\/} in five
bands from 3.6--24\micron\ of the northern inner radio lobe of Centaurus~A, the
nearest powerful radio galaxy.
We show that this emission is synchrotron in
origin.  Comparison with ultraviolet observations from {\it GALEX\/} shows that diffuse
ultraviolet emission exists in a smaller region than the infrared but also
coincides with the radio jet. 
We discuss the possibility, that synchrotron emission is responsible for the ultraviolet emission and conclude that further data are required to confirm this.
\end{abstract}


\section{Introduction}

The radio source Centaurus~A and its host galaxy, NGC\,5128, provide a rare
opportunity to observe the detailed behaviour of a recently merged system supporting
a powerful active nucleus with jets and extended radio emission.  At a distance of
3.4\,Mpc \cite{israel98}, such that  1\arcs = 16.5\,pc, Centaurus~A is the nearest powerful radio galaxy, and its
activity, merger remnants, and star-formation may be resolved
and studied in detail
.

The host galaxy is believed to be a giant elliptical galaxy that recently merged
($\sim$200\,Myrs ago) with a small spiral galaxy, producing the prominent dust lanes
seen in optical images  \citep{bm54,QGF93}.  The active nucleus gives rise to a jet
and counter-jet
 in the inner
arcminute ($1\arcm \approx 1$\,kpc) about the nucleus
\citep[][]{feigelson81,hardcastle03}.  On larger scales, giant radio lobes extend
over $\sim$6\deg\ on the sky, with inner radio lobes extending about 6\arcm\ to the
north-east and south-west.  The focus of this {\em Letter} is the Northern radio `jet', \squig~3--4\,kpc from the nucleus (see Figure 1).  For a
comprehensive review of Centaurus~A see, for example, Israel~(1998), Morganti et al.~(1999) and Junkes et al.~(1993).

While jet/lobe emission is broadly understood in terms of synchrotron emission
from electrons accelerated in the nuclear jet or in associated shocks, there remain uncertainties when interpreting the details of observed
emission.  Observing the synchrotron spectrum, particularly at sufficiently high
frequencies that the cut-off due to spectral aging is measured, allows the study of the energy distribution of the underlying electron population.

Since the radio emission is produced by long-lived electrons, it alone does
not provide detailed information about the underlying physical structure of the
emitting plasma.  At sites of acceleration, highly energetic electrons may
emit at frequencies as high as X-rays \citep[][]{smith83,feigelson81}, so a
multi-frequency approach is required to study these regions.  Infrared (IR) observations
provide important constraints at intermediate frequencies in modelling targets.

The {\it Spitzer Space Telescope\/} \citep[][]{werner04} offers a powerful new capability to study
IR emission from jets in \hbox{AGN}.  The Infrared Array Camera
\citep[IRAC;][]{fazio04} and the Multiband Imaging Photometer
\citep[MIPS;][]{rieke04} provide imaging at 3.6 to 160\microns, with sensitivity orders of magnitude better than any other telescope. 
Whilst  near-IR detections of FRII sources have been made recently (e.g. Floyd et al.~2006),
 no jet to date has
been observed extensively in the IR. 

The {\it Spitzer Space Telescope\/}'s advantage, in addition to offering several IR
bands, is its sensitivity. 
In this {\it Letter} we present IRAC and
MIPS photometry of the northern radio lobe, based upon  effective
integrations of 72\,s and 160\,s respectively, and use them to describe and
interpret the jet SED of Centaurus \hbox{A}.  The IRAC observations and general
processing are described in 
Quillen et al.\ (2006).  The MIPS data are presented here for
the first time.  We use radio data at 843\,MHz, 1.4\,GHz, and 4.9\,GHz
from the literature:  the 843\,MHz data are taken from the Sydney University Molonglo
Sky Survey \citep[SUMSS;][]{sumss};  the 1.4\,GHz and 4.9\,GHz data are from 
Condon et al.\ (1996) and 
Burns et al.\ (1983), respectively.   We show that the diffuse IR
emission is also synchrotron in origin, and compare it to ultraviolet (UV) emission detected by the {\it Galaxy Evolution Explorer\/}
\citep[{\it GALEX\/};][]{martin05}.   \S\,\ref{observations} presents the MIPS imaging and
describes the IRAC, UV, and radio data sets.  \S\,\ref{measure} describes
the method for measuring the surface brightness of the jet and presents the derived
photometry.  \S\,\ref{discussion} interprets the results in terms 
of an underlying synchrotron spectrum and discusses the physical implications.

\section{Observations}
\label{observations}

The multi-wavelength emission we describe (Figure 1) coincides with the part of the jet/lobe
system usually referred to as the northern inner radio lobe (NIRL). 
The NIRL lies between the inner jet and the `large-scale jet-like feature' which connects the NIRL to the middle radio lobe (Morganti et al.~1999).
Centaurus~A is a Fanaroff-Riley class I (FR\,I; Fanaroff \& Riley~1974)
radio source, and has neither the hotspots nor the clear distinction between jet and
lobe that characterize FR\,II sources.  
Referring to a part of the jet in Centaurus A as a 
``lobe'', therefore, can be confusing, particularly in the context of this {\it Letter} which discusses its relation to the nucleus. We instead use the  term ``jet'' to describe the
source.
Figure~\ref{mipsirac} shows the MIPS 24\microns\ image overlaid with 1.4\,GHz\ radio contours and should clarify the source position with respect to
the literature.

\subsection{MIPS imaging}

\begin{figure*}
\begin{tabular}{ccc}
\includegraphics[width=6.0cm,angle=0]{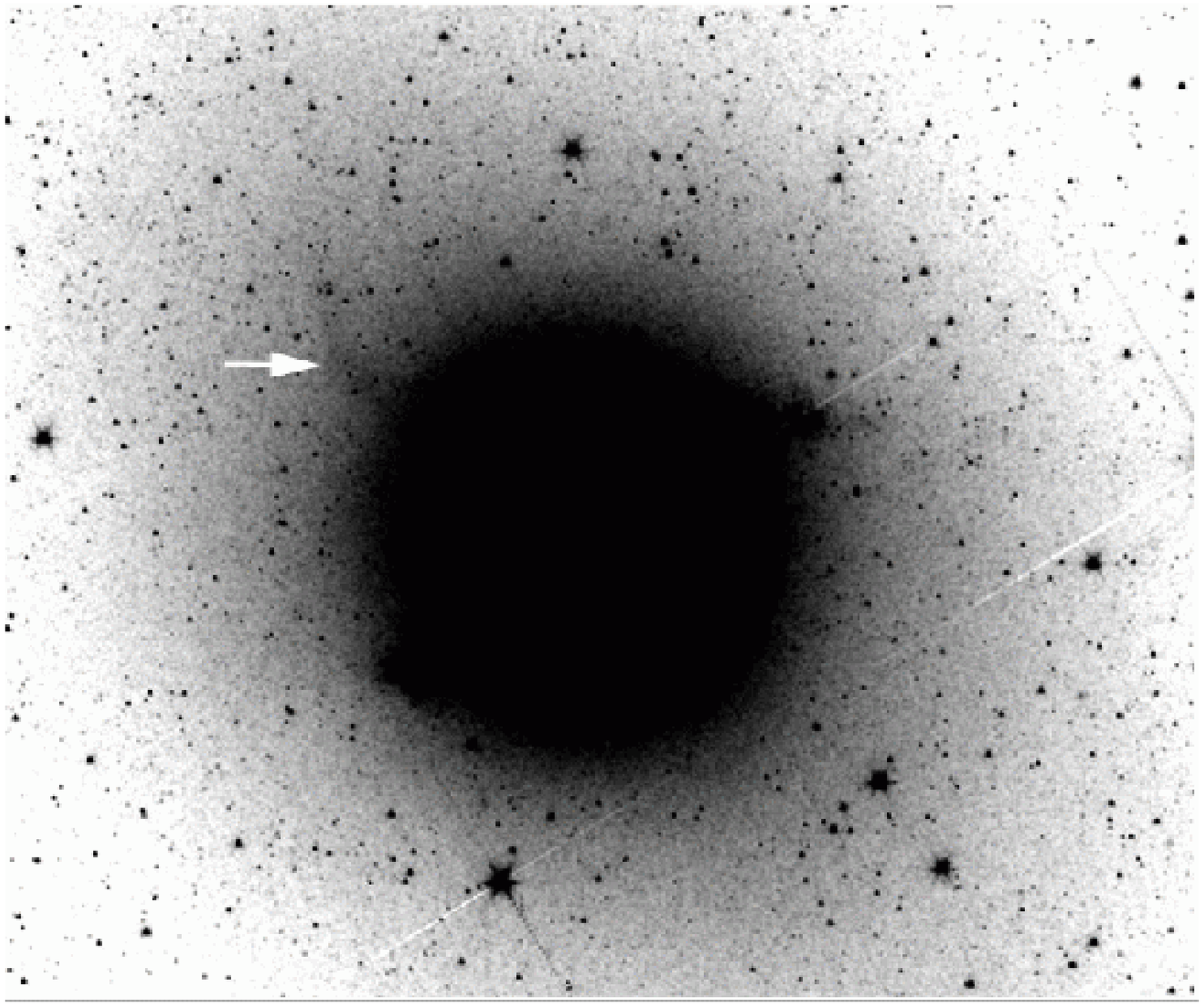}&
\includegraphics[width=6.0cm,angle=0]{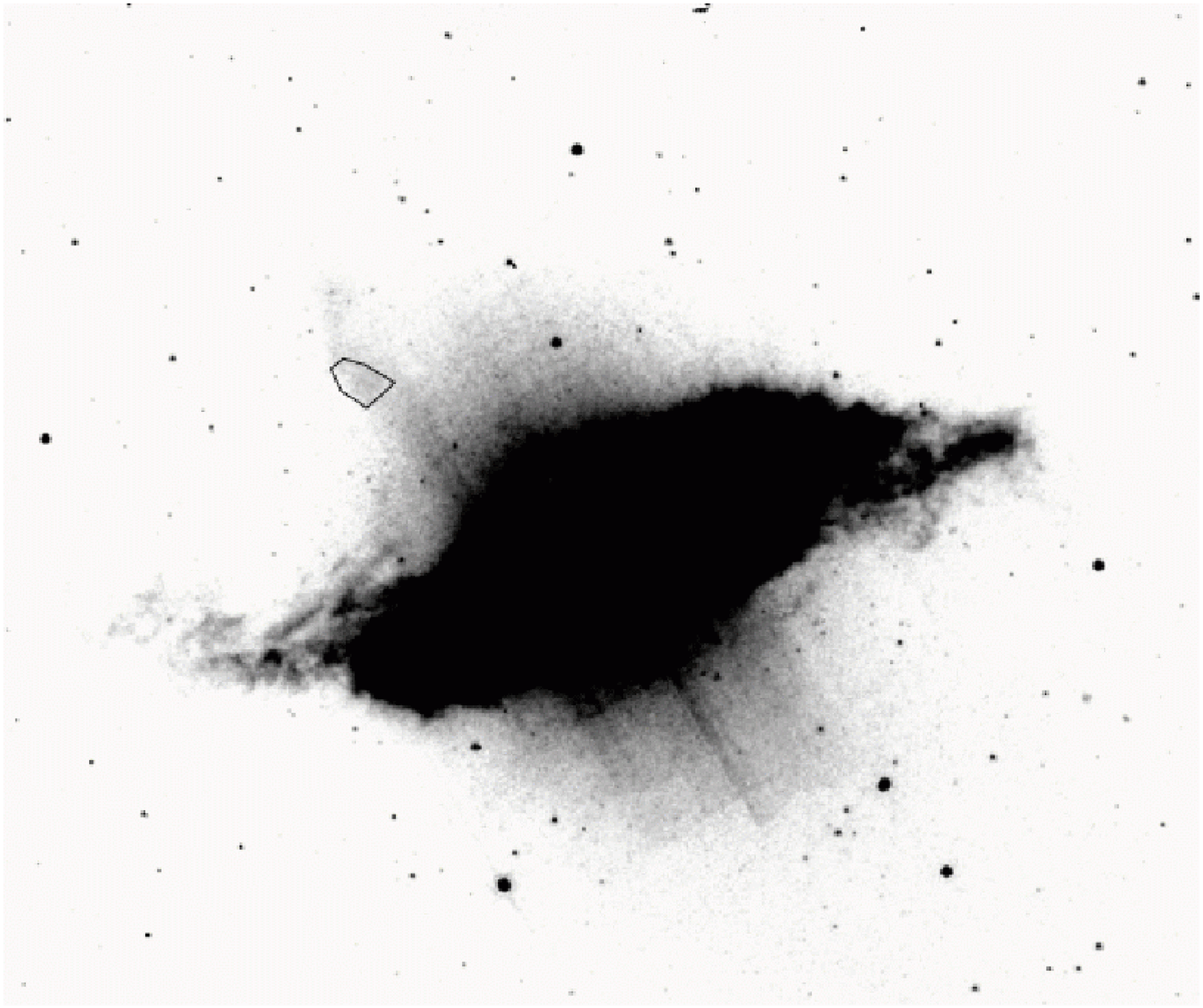}&
\includegraphics[width=5.0cm,angle=90]{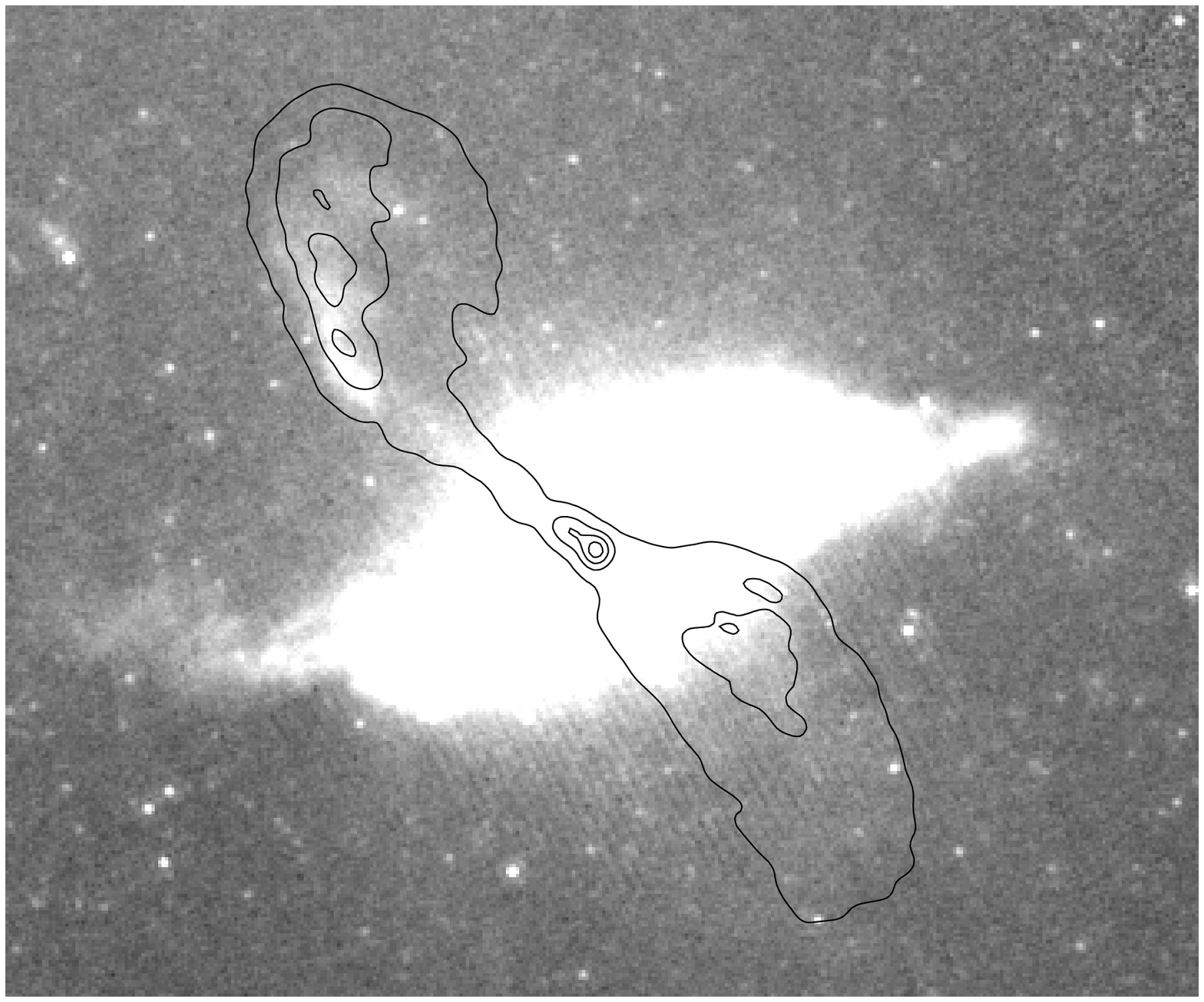}\\
\end{tabular}
\figcaption{Left to right: 3.6\microns\ IRAC image, 8\microns\ IRAC image and
 24\microns\ MIPS image.  
These images show  a 15\farcm2 $\times$ 10\farcm5 portion of the final mosaics. They are centred on 13 25 27.5 -43 1 8.5 (J2000), with North towards the top of the page  and East to the left.
All are stretched to highlight the jet and show the
increasing strength of emission towards longer wavelengths.  The IRAC images show the
aperture used to measure the surface brightness, and the contours on the 24\microns\
MIPS image describe the 1.4\,GHz emission and show how the IR emission coincides
with the radio jet.
\label{mipsirac}
}
\end{figure*}

Centaurus~A was observed with MIPS on 2004 August 6 using scan mapping mode with
14~scan legs at a medium scan rate, resulting in a total exposure of $\sim$160\,s at
the center of the map.  MIPS has three bands centered at 24, 70 and 160\microns\
with bandwidths of 4.7, 19 and 35\microns, respectively. Due to the presence of the bright, extended, dusty disk, measuring the surface brightness of the jet has only been possible at 24\micron\ and we present only that data here.
The 24\micron\ band ``jail bar'' artifacts were corrected by dividing each Basic
Calibration Data (BCD) frame by a normalized median frame (based on all BCDs
excluding the source).  These corrected BCDs were then mosaiced using the
MOPEX  software using single,
multi-frame, and dual outlier rejection (http://ssc.spitzer.caltech.edu/postbcd/).
The final mosaiced image  is $\sim21\arcm \times 
56\arcm$ and  reaches a sensitivity of 0.038\,MJy\,sr$^{-1}$, with a  pixel scale of 2\parcs5; the resolution is $\sim$6\arcs\ (FWHM).
Figure \ref{mipsirac} (far right) shows a 15\farcm2 $\times$ 10\farcm5 section centred on the galaxy. 
The contours
describe the 1.4\,GHz radio emission of the inner radio lobes, and show that the the
jet is clearly detected at 24\microns. The counter-jet/southern radio lobe
is not detected.

\subsection{Data at other wavelengths}

IRAC observations at 3.6, 4.5, 5.8, and 8\microns\ are taken from Quillen et
al.~(2006) and the 3.6\microns\ and 8\microns\ images are shown in Figure~1 (left
and middle panels), stretched to emphasize detection of IR emission coincident
with the radio lobe.   These mosaiced observations had a typical exposure time of
72\,s at the source, and reached depths of 0.025, 0.024, 0.07 and \hbox{0.06 MJy
sr$^{-1}$} in bands 1--4, respectively.  Radio data at 1.4\,GHz from Condon et
al.~(1996) show the jet with good signal-to-noise ratio, as do 843\,MHz data
taken from SUMSS \citep[][]{sumss}.   In addition, a re-reduced version of the 4.9\,GHz data \citep[][]{burns83} was provided by M. Hardcastle (priv.\ comm.) and used to put limits on the radio emission at that frequency.   Ultraviolet observations obtained with {\it GALEX\/}
\citep[][Neff et al. in prep.]{neff03} at near and far UV  bands cover 1350--1800\AA\ and 1800--2800\AA.   The effective
wavelengths for the bands are 1528 and 2271\AA respectively.

\section{Surface Brightness measurements of the Northern Inner Radio Lobe}
\label{measure}

\begin{figure*}
\plotone{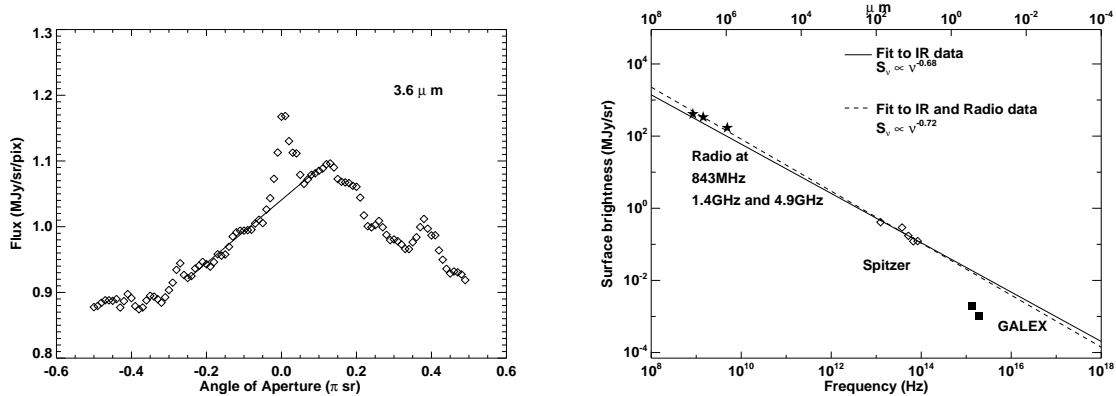}
\figcaption{Left: Mean counts per aperture as a function of angular position about
the nucleus.
The abscissa zero point corresponds to the original aperture, chosen by eye to fit the jet shape.
The jet surface brightness is determined by subtracting an estimate of the background (fitted to surrounding apertures and shown by the solid line) from the total surface brightness.
Right: The spectral energy distribution of the jet of Centaurus~A. 
The surface brightness was determined from a six-points aperture, as described in the text.
Errors on the photometry are smaller than the plotting symbols used.
\label{bgfit_jetsed}
}
\end{figure*}

Measuring the surface brightness of the jet requires careful subtraction of the
underlying emission from the host galaxy.
This was done by creating an irregularly shaped aperture matched to the jet, and rotating it about the center of the galaxy
(i.e., the unresolved nucleus in the 4.5\micron\ image).   The surface brightnesses
measured when the aperture was adjacent to the lobe are used to estimate the
background due to the galaxy.  The aperture is a polygon chosen by eye to match the
shape of the lobe in the near UV {\it GALEX\/} image\footnote{As discussed
in \S\,\ref{discussion}, diffuse emission associated with the lobe in the near
UV image covers a smaller area than the IR and radio emission in the
region.   The aperture is therefore based upon the UV data in order that
the comparison of the mean surface brightness be fair.}.   It covers
653~pixels in the IRAC images (1\parcs22/pixel) and has six points starting at 13 26 24.69 -43 10 16.4 (J2000), with vertices offset by 1.78, 2.47, 1.64, 0.13 and -1.92\,s\ East and 13.5, 30.0, 37.5, 33.0 and 19.5\arcs\ North, respectively.
Figure~\ref{mipsirac} (middle) shows the
position of the jet aperture overlaid on the 8\micron\ IRAC image.  
The aperture was rotated about the center of the galaxy in 100 steps covering
$\pi/2$ radians on either side of the jet.  At each position, the counts in the
aperture were summed, excluding pixels on or near stars or other discrete sources.
The mean count per aperture was then plotted as a function of angular
separation from the radio lobe.  By excluding the region corresponding to the radio
lobe and fitting the mean count from apertures on either side of the lobe, an
estimate of the background at the lobe was made and subtracted.  Linear and
quadratic fits to the galaxy background were made.   
Figure~\ref{bgfit_jetsed}
(left) illustrates the  3.6\,\micron\ measurement. An estimate of the photometric 
error was taken to be the standard deviation of the difference
between the fit to the background and the measured background across the range of
the fit. This typically indicated a 10$\sigma$\, detection and shows that this
measurement error dominates the 5\% photometric error which is typical of IRAC
mosaics. This method was repeated in all four IRAC images, the MIPS 24\micron\
image, the {\it GALEX\/} images, and the radio images. 

\section{Results and Discussion}
\label{discussion}

Figure \ref{bgfit_jetsed} (right) shows the surface brightness of the designated
region of the jet of Centaurus~A as a function of frequency.  The solid
line is a power law, $S\propto \nu^{-\alpha}$, fit to the IR points for which
$\alpha$ = 0.68.   
This line comes remarkably close to the radio points suggesting that the IR emission is also synchrotron in origin, though the radio points are slightly flatter than the IR points. 
This suggests the possibility of a break in a synchrotron spectrum, rather than a simple, single power-law spectrum. However this is not a strong conclusion.
We conclude
that the IR emission is due to synchrotron emission and, when a single power-law is fitted to both the radio and IR points, $\alpha$ = 0.72 (the dashed line in Figure \ref{bgfit_jetsed}).

Synchrotron emitting electrons radiate at a rate proportional to their energy.  The
most energetic electrons lose their energy the fastest, leading to a slope that is
one-half power steeper in the high frequency spectrum,  at frequencies much above the break frequency, $\nu_B = eB/2\pi m_{\rm e}c$ (e.g. Beams and
Jets in Astrophysics, ed. Hughes, 1991). 
Our observations indicate that this break must occur at frequencies higher than
$10^{13}$\,Hz.   Though detailed modelling is required for an exact evaluation of
the break frequency, an approximate limit may be gained.  Following 
Miley (1980),
we calculate the magnitude of the magnetic field on the assumption that the system
is at its minimum energy density.   Using Equation~3 of 
Miley (1980) at 1.4\,GHz,
we find $B\approx3$\,\hbox{nT}.  Using the highest IR frequency observed
(83\,THz), an upper limit to the lifetime of the emitting electrons is found to be 30,000\,yrs.

\begin{figure*}
\plotone{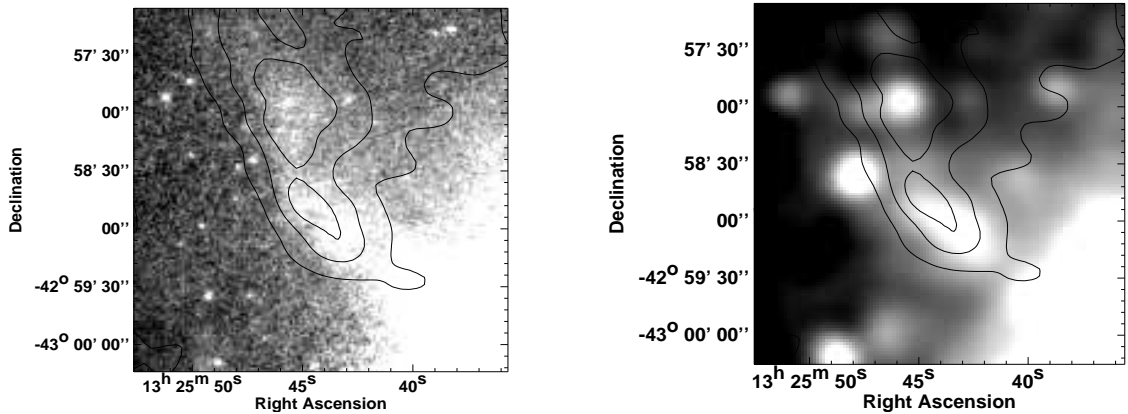}
\figcaption{
Left: 8\microns\ IRAC image of the jet with contours describing the 1.4\,GHz radio
emission. Right: Near UV {\it GALEX\/} image (smoothed by a 7\farcs6 circular Gaussian) with 1.4\,GHz radio contours.  The two sources to the North and East are discrete and are likely to be associated with star-formation sites.
Both images show the region of the aperture used to measure the surface brightness (Figure \ref{mipsirac}).
\vspace*{1.0cm}
\label{overlayuv_plot}
}
\end{figure*}

It is important to know how far the synchrotron spectrum extends without a
break, since very high frequency emission is produced by short-lived, high
energy particles and may imply in situ particle acceleration.
In both the near UV and far UV {\it GALEX\/} images there are two
kinds of emission: extended sources, likely associated with
star-formation sites; and diffuse emission which coincides with the jet.
Figure \ref{overlayuv_plot} shows the 1.4\,GHz radio contours overlaid on the  8\micron\ IRAC image in greyscale (left) and  the near
UV {\it GALEX\/} image (right; smoothed with a 7\farcs6 circular Gaussian for clarity).  
The UV emission is seen
only at the inner portion of the jet, whereas the IR emission follows the
jet to the north, consistent with the idea that emission at all three
wavelengths is synchrotron emission.  The synchrotron lifetime of the
UV emitting electrons is shorter than that of the IR emitting
electrons, and so they do not live/emit long enough to trace the 
bulk motion downstream.  
northward at this point, indicative of interactions with the environment in the
region.

The surface brightness of the diffuse UV emission is compared to that
of the radio and IR emission in Figure
\ref{bgfit_jetsed} (right).  The UV points are significantly below the extrapolation
of the synchrotron emission based upon the radio and IR data.  Several
explanations must be considered.  
Firstly, the diffuse UV emission might not
be synchrotron at all.  
We consider this unlikely because the jet emission extends  to X-ray
frequencies \citep[][]{kraft03}, though it is possible that other mechanisms, such as diffuse star-formation, also contribute to the UV. 
Secondly, we could be seeing the
effects of extinction due to dust.  
Suppose that the best-fit power law to the radio and infrared
measurements extends without break to ultraviolet frequencies, but that there
is intervening dust.  Assuming a Calzetti extinction law, only $E(B-V) = 0.56$
is required to produce  the observed ultraviolet emission.
Given the violent dynamic history of Centaurus~A there is clearly a
large presence of gas, and therefore dust, associated with the merger, which can be found in distinct regions far outside the galactic disk.
Hence it is plausible that  sufficient extinction, at the shortest wavelengths, exists along specific lines of sight.
Note that  if this were the case, the appropriate synchrotron lifetime for the UV emitting particles (\squig~~6000 years) would be short enough to require {\it in situ} particle acceleration, given  a projected distance of $\sim$3\,kpc from the nucleus, and bulk motion $\sim$0.5\,c in the kpc scale jet \citep{hardcastle03}.

However, inclusion of X-ray measurements argues for a third possibility:
a spectral break between infrared and ultraviolet frequencies.  Hardcastle et
al.\ (2006; henceforth H06) fit a broken power-law spectrum to all measurements from radio to
X-ray frequencies, and determine that a break occurs at $\sim$0.3\,THz.  
Figure 5 of H06 shows that
this broken power law accounts well for the
radio, infrared, and X-ray measurements.  However, it significantly 
underpredicts the ultraviolet measurements, even assuming no extinction from
dust intrinsic to Centaurus A (Galactic extinction was included).  If intrinsic
extinction is important this discrepancy becomes worse.  

Neither dust alone, nor a broken power-law accounts for the data in an
entirely satisfactory way. Submillimeter and 70\micron\ {\it Spitzer\/}
measurements with higher SNR would fill in the gap in the observed 
spectrum above and below 1\,THz, providing significant new constraints on the
overall spectrum, and would help in resolving the issue.


\section{Summary}

New IR observations of the jet, at the position of the NIRL associated with Centaurus~A obtained
with the {\it Spitzer Space Telescope} show that synchrotron emission extends at
least from radio to IR wavelengths.  Diffuse UV emission is also
present in the jet and it is plausible that this emission is also synchrotron in
origin, reddened by intervening dust.  
X-ray data imply a break in the spectrum, but further data are required to constrain this precisely.



\acknowledgments

This work is based on observations with the NASA {\it Spitzer Space Telescope\/}, which is operated by the California Institute of Technology for NASA under NASA contract 1407.
We wish to thank Susan Neff from the {\it GALEX\/} team for providing the UV data and her useful discussions.
We also thank Martin Hardcastle for providing the re-reduced 4.9\,GHz radio data and his part in coordinating the publication of these results.

\clearpage

\end{document}